# Mesoscale Imperfections in MoS$_2$ Atomic Layers Grown by Vapor Transport Technique

Yingnan Liu[1*], Rudresh Ghosh[2*], Di Wu[1], Ariel Ismach[2], Rodney Ruoff[2,3], Keji Lai[1,**]

[1]Department of Physics, University of Texas at Austin, Austin, TX 78712, USA

[2]Department of Mechanical Engineering, University of Texas at Austin, Austin, TX 78712, USA

[3]IBS Center for Multidimensional Carbon Materials, Ulsan National Institute of Science and Technology (UNIST), Ulsan 689-798, Republic of Korea

[*]These authors contribute equally to the paper.

[**]Address correspondence to kejilai@physics.utexas.edu

## Abstract

**The success of isolating small flakes of atomically thin layers through mechanical exfoliation has triggered enormous research interest in graphene and other two-dimensional materials. For device applications, however, controlled large-area synthesis of highly crystalline monolayers with a low density of electronically active defects is imperative. Here, we demonstrate the electrical imaging of dendritic ad-layers and grain boundaries in monolayer molybdenum disulfide (MoS$_2$) grown by vapor transport technique using microwave impedance microscopy. The micrometer-sized precipitates in our films, which appear as a second layer of MoS$_2$ in conventional height and optical measurements, show ~2 orders of magnitude higher conductivity than that of the single layer. The zigzag grain boundaries, on the other hand, are shown to be more resistive than the crystalline grains, consistent with previous studies. Our ability to map the local electrical properties in a rapid and nondestructive manner is highly desirable for optimizing the growth process of large-scale MoS$_2$ atomic layers.**

Single-layer molybdenum disulfide ($MoS_2$) is an atomically thin direct band gap ($E_g$ = 1.8eV) semiconductor [1, 2]. Followed by the discovery of thickness dependent indirect-to-direct-gap transition [1, 2] and valley quantum physics [3–6], many device prototypes, such as field-effect transistors [7–10], photo detectors [11, 12], and floating gate memories [13], were reported on exfoliated $MoS_2$ samples. In order to meet the needs for next generation nanoelectronics, however, the production of large-area and high-quality monolayer $MoS_2$ films is a must. In the past few years, significant effort has been devoted to this research direction and the progress has been breathtaking [14–20].

The amount of imperfections on the chemically grown monolayer $MoS_2$ is a good measure of the film quality. Point defects, including vacancies of the host atoms and unintentional impurity atoms, usually receive the most attention because they directly dope the semiconductor and shift its chemical potential [21, 22]. There exist, on the other hand, defects in the mesoscopic (nanometer to micrometer) length scale. For example, several experimental and theoretical studies have investigated the influence of grain boundaries (GBs) on the local electronic properties [19–21]. Moreover, it is widely observed that micrometer-sized irregular ad-layers are overgrown on the $MoS_2$ monolayer [19, 20]. If their electrical properties deviate significantly from the underlying single layers, these mesoscale imperfections could dominate the performance when the device dimension approaches the same length scale. It is thus crucial to perform spatially resolved electrical characterization, preferably in a rapid and nondestructive fashion, on various as-grown $MoS_2$ thin films.

In this study, the $MoS_2$ atomic films on $SiO_2$ (285nm)/Si substrates were prepared by the sulfurization of $MoO_3$, a process similar to those described in references [15, 19, 20] (see Methods and Supporting Information S1 for details). Figure 1 shows the scanning electron microscopy (SEM) images of the deposited film at various distances away from the $MoO_3$ precursor. At the onset of monolayer growth, small isolated $MoS_2$ islands that are nearly round in shape are observed in Fig. 1(a). Moving towards the $MoO_3$ source, the flakes become bigger in size and more triangular in shape [Fig. 1(b)]. As the size of individual triangles reaches 30~50μm, neighboring grains start to merge [Fig. 1(c)] and eventually form a pseudocontinuous film [Fig. 1(d)]. From the SEM images, two types of mesoscale defects are clearly seen on the monolayer $MoS_2$. As shown in Fig. 1(e), dendritic precipitates with various sizes are randomly distributed within individual grains. The merging of different islands is more complicated. As previously reported [20], the ad-layer overgrowth may or may not occur on the grain boundaries, both observed in our samples [Fig. 1(f) and Fig. 1(g)]. While our current effort in growth aims to minimize the areal density of these defects, the goal of this work is to fully evaluate their electrical properties with respect to the underlying monolayer $MoS_2$.

Figs. 2(a-c) show the conventional Raman and photoluminescence (PL) studies of a single domain monolayer $MoS_2$ (more details in Supporting Information S2). Due to its sensitivity to inter-layer coupling, the well-established Raman spectroscopy has become the standard technique to determine the number of layers in 2D van der Waals materials [23]. At the same time, the direct band gap in monolayer $MoS_2$ leads to a pronounced PL peak [1] when excited by the same laser in our Raman setup. As seen in the Raman and PL maps, the dendritic regions display a wider $E^1_{2g}$-$A_{1g}$ frequency separation and a much weaker PL signal than the monolayer $MoS_2$ background. Combined with the atomic force microscopy [AFM, Fig. 2(d)] data, all conventional measurements suggest that the irregular ad-layers are structurally the same as $MoS_2$ bilayers.

An AFM-based microwave impedance microscope (MIM) was employed to study the local conductivity of the synthesized $MoS_2$ atomic layers. MIM is the GHz-frequency analogue of near-field scanning optical microscopy (NSOM) and measures the local electronic rather than optical responses [24–26]. Compared with DC probes, however, microwave frequency is high enough so that no contact electrode is needed for the imaging, therefore nondestructive to the sample. This unique capability allows us to rapidly map out the electrical properties of many dendritic features for statistical analysis. During the measurements, an excitation signal at 1GHz is delivered to the metallic tip of a special cantilever probe through the electrically shielded center conductor [27]. The local variation of the sample permittivity and conductivity modifies the effective tip impedance, which results in small changes of the reflected microwave [28]. The RF electronics then detect such signals to form the real (MIM-Re) and imaginary (MIM-Im) components of the MIM output. Details of the MIM setup are listed in the Methods and Supporting Information S3.

Figs. 2(d) and 2(e) show the simultaneously taken AFM and MIM-Im images of the same flake measured by the Raman [Fig. 2(a)] and PL [Fig. 2(b)] mapping. Strikingly, while the AFM again measures double layer thickness in the dendritic regions, the corresponding MIM signals are substantially larger than that on the monolayer MoS$_2$. Moreover, the line cut in Fig. 2(f) shows that bigger dendrites display stronger MIM-Im signals. The results are consistently observed across the substrate, regardless of the size of the monolayer islands [Fig. 3(a, b)]. Fig. 3(c) summarizes the peak MIM-Im and MIM-Re signals of individual dendritic patches, showing a roughly linear areal dependence of the signal strength. For comparison, the MIM signals on the monolayer itself, plotted on the same graph, is barely discernible over the SiO$_2$/Si substrate and essentially independent of the domain sizes. Interestingly, on a small number of MoS$_2$ samples, a triangular island with one extra layer thickness can be found in the middle of the underlying monolayer domain. As shown in Supporting Information S4, the MIM signals on these regularly shaped bilayers are much closer to the monolayers than the dendritic ad-layers. We therefore conclude that, in terms of electrical properties, the irregular dendrites are very different from triangular bilayers with good crystallinity.

For a quantitative understanding of the dendritic precipitates and monolayer MoS$_2$ in our films, we have performed the finite-element analysis (FEA) [28] using a numerical software COMSOL4.3 and the result is shown in Fig. 3(d). The simulation takes the experimental geometry and other known parameters, e.g., the permittivity of SiO$_2$ and monolayer MoS$_2$ [29, 30], to calculate the effective tip impedance. In this case, the MIM-Re and MIM-Im signals are proportional to the real and imaginary parts of the change of tip admittance (reciprocal of impedance), which allows a direct comparison between the experimental data and the simulation results. Due to the scattered data points and complicated nature of the near-field interaction, it is difficult to accurately quantify the sample conductivity ($\sigma$). However, as detailed in Supporting Information S5, using reasonable assumptions of the tip geometry and tip-sample contact condition, an order of magnitude estimate of the local conductivity is feasible within the measurement and statistical error. For the monolayer MoS$_2$ flakes, the modeling yields a low $\sigma$ of 1~10 S/m or a sheet resistance of $10^8$~$10^9$ $\Omega$. The MIM data on the dendritic ad-layers, however, suggest a much higher $\sigma$ of $10^2$~$10^3$ S/m. This value, while still much lower than that of a typical metal, implies a substantial amount of carriers in the ad-layers. Due to the relatively high local conductivity, the quasi-static potential drop within the ad-layer is small, which explains the fact that the MIM signals roughly scale with the areas (Supporting Information S5). Since the lateral dimensions of these mesoscale patches are comparable to the sizes of typical nano-devices, their effects are certainly not negligible for electronic applications. Much more work is needed to determine whether similar dendritic regions in other CVD-grown films have the same conductivity as reported here.

Finally, we briefly discuss the MIM imaging of $MoS_2$ grain boundaries. Theoretical and experimental work has been carried out to understand the atomic structure and electronic property of various GBs in $MoS_2$ monolayers [19–21]. In our samples, most GBs are covered by a second layer, which complicates the interpretation of MIM data in terms of the underlying GBs. We did, however, observe some flat GBs without the overgrown layer. As seen in Fig. 4, the zigzag GB appears darker than the grains in the MIM-Re data. Although the GB width is obviously broadened due to our limited spatial resolution, the result suggests that the local conductivity near the zigzag GB can be one order of magnitude lower than the background, consistent with previous transport and DFT calculations [19]. Because of the finite resolution of the MIM imaging, either the GB itself or defect segregation around the GB could contribute to the observed high local resistance.

In summary, we report the nanoscale impedance microscopy of two types of mesoscopic electrical imperfections in $MoS_2$ atomic layers grown by vapor transport technique. The micrometer-sized dendritic ad-layers, being identical to bilayer $MoS_2$ from conventional structural and optical measurements, display ~2 orders of magnitude higher conductivity than that of the monolayer. The zigzag grain boundaries, on the other hand, lower the local conductivity of $MoS_2$. Compared with transport measurements, the MIM provides direct nanoscale electronic information without the need for fabricating ohmic contacts or other cumbersome sample preparations. We suggest that such capability will be particularly useful for the fast growing field of 2D van der Waals materials.

**METHODS**

**CVD Growth of $MoS_2$:** The $MoS_2$ atomic layer films were grown by standard vapor transfer growth process. The inner diameter of the quartz tube is 25 mm. The starting materials were $MoO_3$ (15 mg) and sulfur (1g) powder that were loaded in alumina crucibles and placed inside the tube. The substrates (285nm $SiO_2$ on Si) were placed with the polished side facing the $MoO_3$ powder. The tube was then sealed and evacuated to a base pressure of 10 mTorr. After the base pressure was reached, the vacuum pump was shut off and $N_2$ gas was introduced into the tube at a flow rate of 200 sccm. Once the atmospheric pressure was reached, the $N_2$ gas flow was reduced to 10 sccm. At this point, the temperature of the furnace was raised to 850°C at a rate of 50°C/minute. The temperature of sulfur at one end of the furnace is roughly 350°C. The growth continued for 5 minutes at 850°C, after which the heater in the furnace was turned off and the $N_2$ flow rate was increased to 200 sccm for cooling down.

**Raman and Photoluminescence Measurements**: Raman and PL spectroscopy was done using a Witec Alpha 300 micro-Raman confocal microscope, with the laser operating at wavelength of 488 nm. Parameters in our mapping were (i) grating (Raman) = 1800g/mm, (PL) = 600g/mm; (ii) integration time = 1 second; (iii) resolution = 3 pixels/micron.

**Microwave Impedance Microscopy Measurements:** The MIM in this work is based on a standard AFM platform (ParkAFM XE-70). The customized shielded cantilevers are commercially available from PrimeNano Inc. Finite-element analysis was performed using the commercial software COMSOL4.3. Details of the MIM experiments and numerical modeling can be found in the Supporting Information S3 and S5.

*Conflict of Interest*: The authors declare no competing financial interest.

*Acknowledgment*. The MIM work (Y.L., D.W., K.L.) was supported by Welch Foundation Grant F-1814, and the $MoS_2$ synthesis (R.G., A.I., R.S.R.) was supported the Center for Low Energy Systems Technology (LEAST), one of the six centers supported by the STARnet phase of the Focus Center Research Program (FCRP), a Semiconductor Research Corporation program sponsored by MARCO and DARPA. The authors appreciate helpful discussions with Prof. Chih-Kang Shih and Prof. Deji Akinwande.

*Supporting Information Available:* Details of the growth chamber, Raman and PL data, MIM electronics, comparison between ad-layers and bilayers, and FEA simulation. This material is available free of charge via the Internet at http://pubs.acs.org.

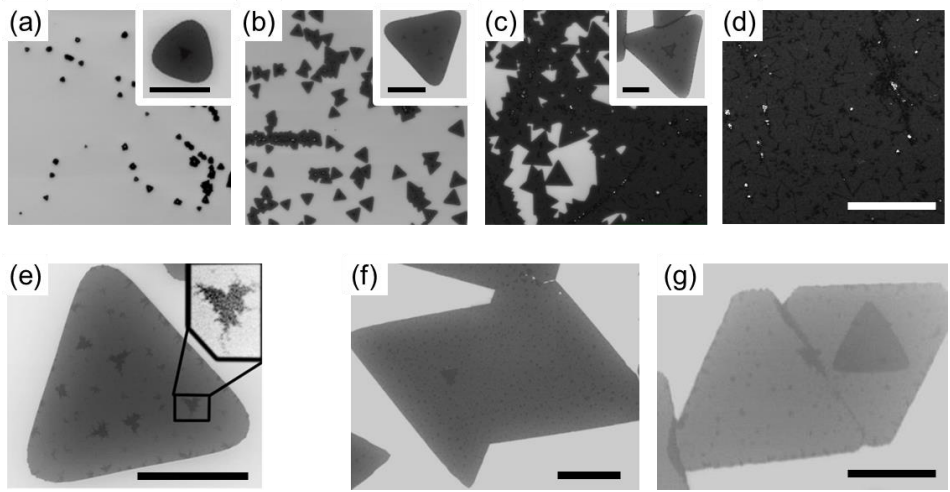

Figure 1: (a-d) SEM pictures of the CVD grown MoS$_2$ film as a function of distance from far to near the MoO$_3$ precursor. The insets show typical individual MoS$_2$ domains with different shapes and sizes. (e) A close-up view of the dendritic ad-layers, which are randomly dispersed on the surface of monolayer MoS$_2$. When two domains of MoS$_2$ merge, the grain boundary (f) may or (g) may not show an overgrown layer. All scale bars in black are 20μm in length and the white scale bar in (d) is 100μm.

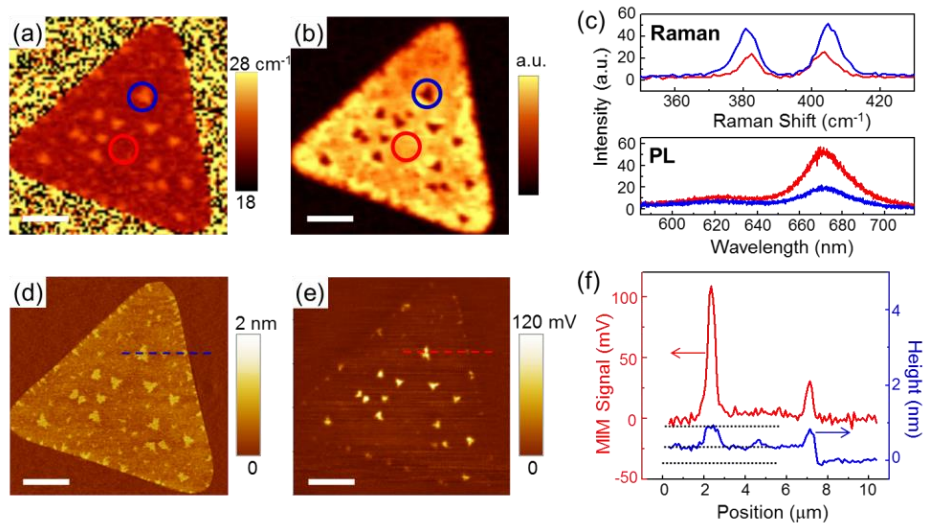

Figure 2: (a) Raman map of the separation between the $A_{1g}$ and $E^1_{2g}$ peaks of one $MoS_2$ flake. (B) Map of the photoluminescence intensity on the same $MoS_2$ domain. (c) Raman and PL spectra for both monolayer (red) and ad-layer (blue) regions circled in (a) and (b). (d) AFM and (e) MIM-Im images of the same domain. (f) Line cuts of the height (blue) and MIM-Im signal (red) in (d) and (e). All scale bars are 5μm.

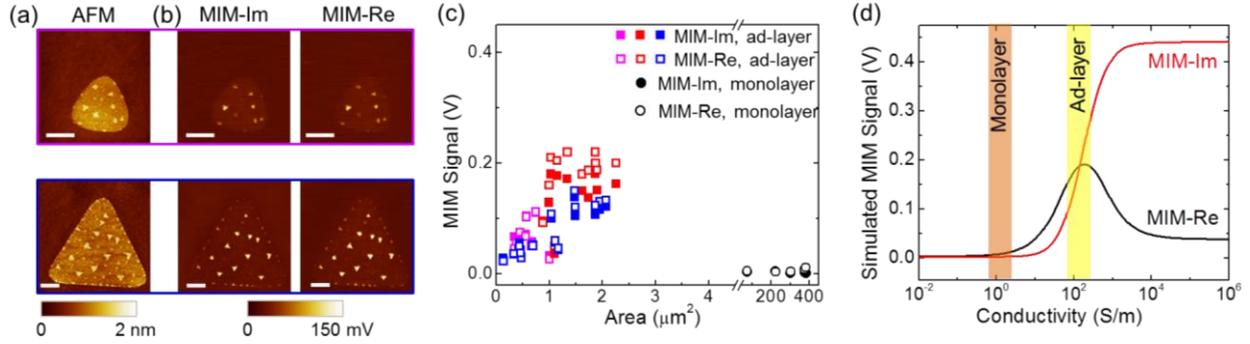

Figure 3: (a) and (b) AFM and MIM-Im/Re images of two MoS$_2$ single domains. (c) Microwave signals of multiple [in different colors, data in purple and blue are from (a)] MoS$_2$ flakes as a function of the area of the ad-layers (squares) and monolayers (circles). All solid symbols represent the MIM-Im signals and empty symbols represent MIM-Re. (d) FEA simulation result of the MIM signals as a function of the ad-layer conductivity. The area of the dendritic region in the modeling is 2μm$^2$. The MIM signals are independent of the area for low sample conductivity and scale with the ad-layer sizes for high σ. All scale bars are 5μm.

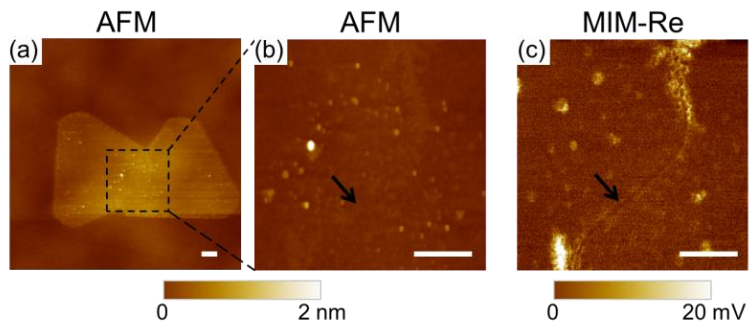

Figure 4: (a) AFM image of two merged monolayer MoS$_2$ domains and (b) a close-up view inside the dashed square. (c) MIM-Re image of the same region in (b). The black arrows indicate the grain boundary, which appears in (c) as a dark curve but invisible in (b). All scale bars are 1μm.

# Supporting Information

# Mesoscale Imperfections in MoS$_2$ Atomic Layers Grown by Vapor Transport Technique

Yingnan Liu[1*], Rudresh Ghosh[2*], Di Wu[1], Ariel Ismach[2], Rodney Ruoff[2,3], Keji Lai[1,**]

[1]Department of Physics, University of Texas at Austin, Austin, TX 78712, USA

[2]Department of Mechanical Engineering, University of Texas at Austin, Austin, TX 78712, USA

[3]IBS Center for Multidimensional Carbon Materials, Ulsan National Institute of Science and Technology (UNIST), Ulsan 689-798, Republic of Korea

[*]These authors contribute equally to the paper.

[**]Address correspondence to kejilai@physics.utexas.edu

## S1: Various growth mechanisms on the substrate

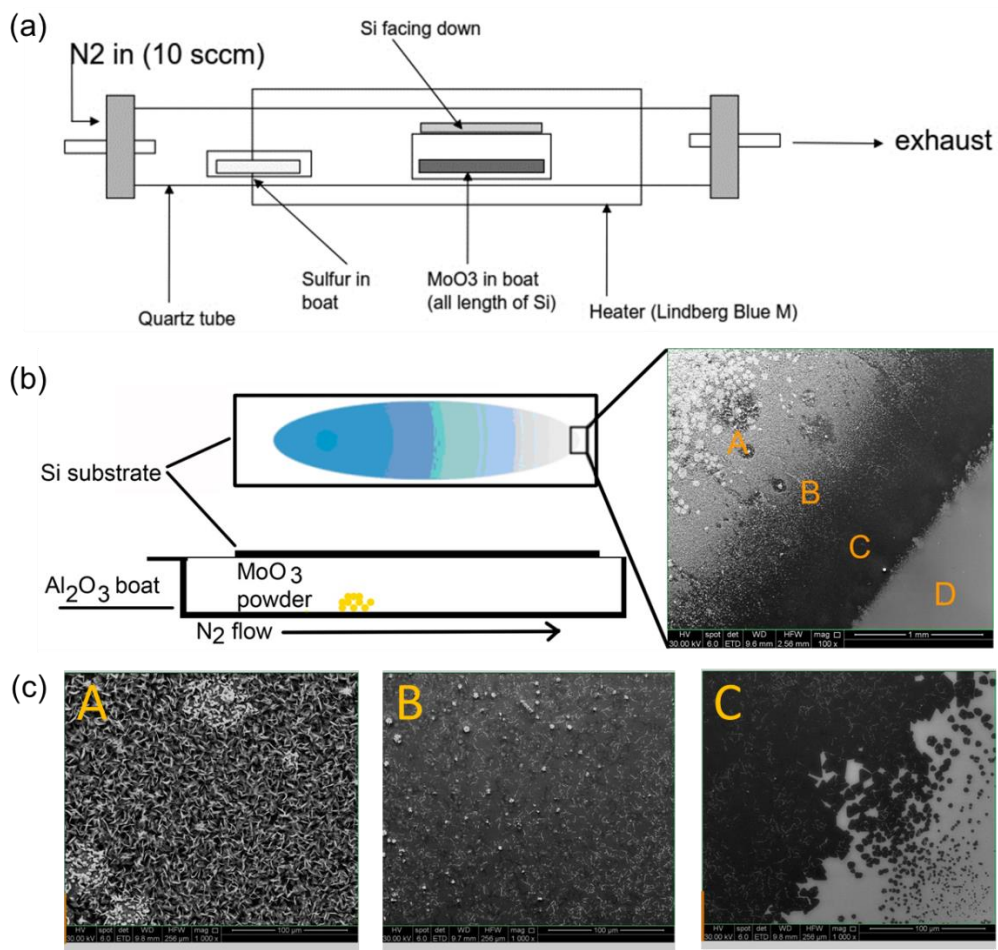

Figure S1(a) illustrates the vapor transfer growth of our $MoS_2$ atomic layer films. The process is detailed in the METHODS section. A schematic of the $MoS_2$ deposition on the $SiO_2/Si$ substrate is shown in Fig. S1(b). Different types of growth mechanisms are observed as a function of the distance from the $MoO_3$ precursor. Close-up SEM images are shown in Fig. S1(c). In region A, bulk-like 3D growth occurs near the source. The deposited $MoS_2$ gradually turns to a pseudo-continuous film from B to C. Isolated monolayer islands appear further away from the $MoO_3$ source, after which only bare $SiO_2/Si$ (region D) is seen.

## S2: Additional Raman and photoluminescence images

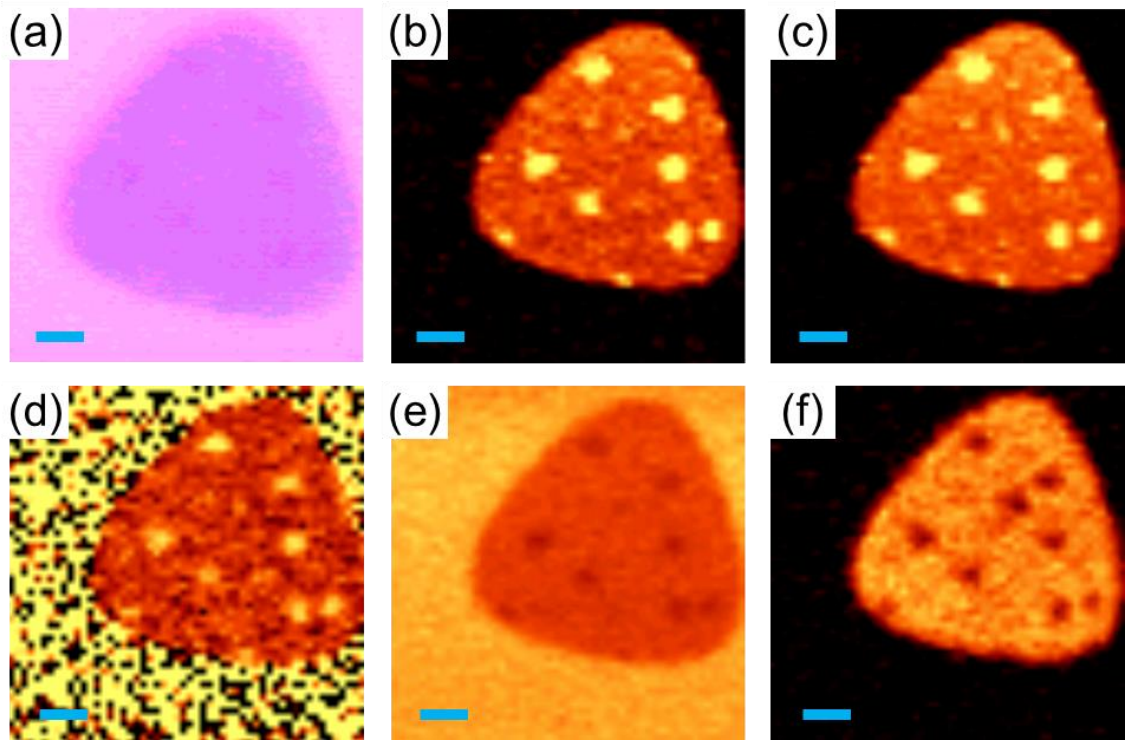

Figure S2 shows a typical mapping set that includes the following frames: (a) Optical microscopy image; (b) Intensity map of $E^1_{2g}$ (sum of counts between 365 and 387 cm$^{-1}$); (c) Intensity map of $A_{1g}$ (sum of counts between 388 and 420 cm$^{-1}$); (d) Map of distance between the two peak positions; (e) Intensity map of the Si peak (sum of counts between 510 and 530 cm$^{-1}$), (f) Intensity map of the PL spectra. All scale bars are 2μm.

## S3: MIM setup

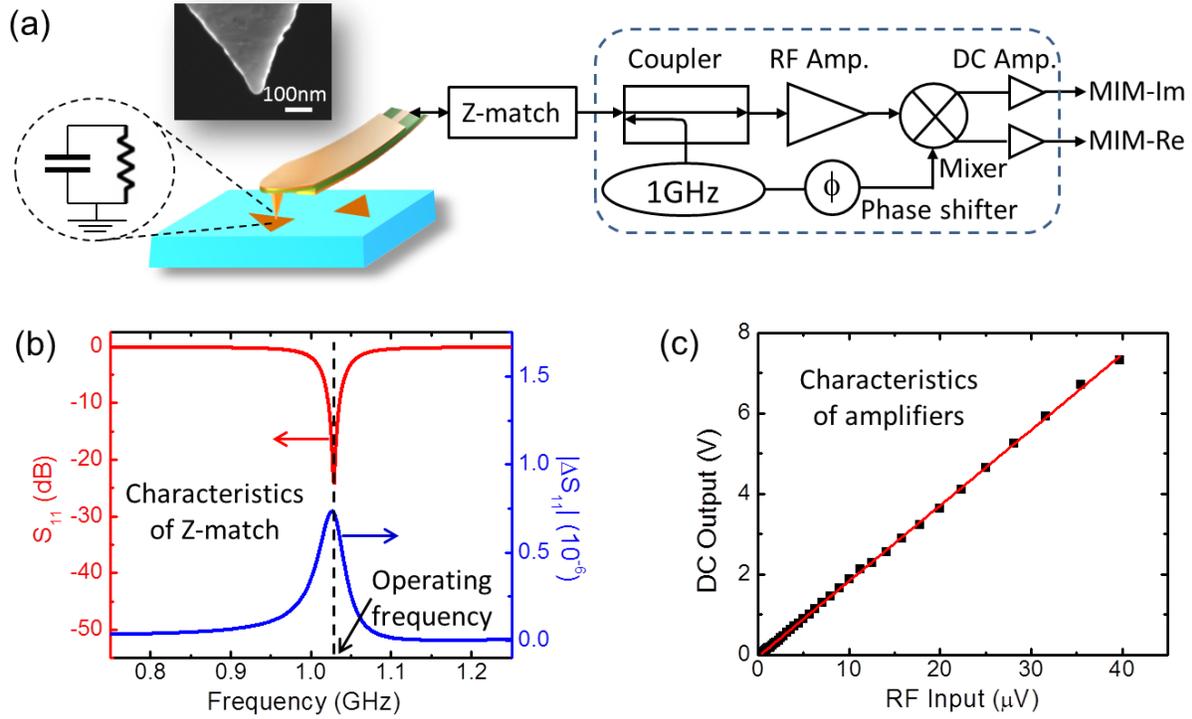

Figure S3(a) shows a schematic of the MIM setup. Details of the shielded cantilever (tip apex shown in the inset) can be found in Ref. [1]. The 1GHz excitation $V_{1GHz}$ ~ 20mV (10μW signal in the 50Ω transmission lines) is guided to the cantilever through a Z-match section [2], which converts the tip impedance (C ~ 1.2pF and R ~ 4Ω) to 50Ω. The reflection coefficient $S_{11}$ from the Z-match section is shown as the red curve in Fig. S3(b). The electronics operate at the frequency of minimum $S_{11}$, which also corresponds to the maximum $\Delta S_{11}$. During the scanning, the near-field interaction changes the effective impedance of the tip, which results a change of the reflected microwave voltage $\Delta S_{11} \cdot V_{1GHz}$. The blue curve in Fig. S3(b) is a calculated frequency-dependent $\Delta S_{11}$, assuming an admittance change of 1nS. The MIM circuit then amplifies this RF input by the RF amplifier, demodulates the RF signal, and further amplifies it as the final output. The amplifier response is shown in Fig. S3(c). In the experiment, the phase shifter in front of the mixer is adjusted such that the two orthogonal channels are aligned to the real and imaginary components of the tip-sample admittance, which can be simulated by FEA modeling (Figure S5).

## S4: Comparison between dendritic ad-layer and triangular bilayer MoS$_2$

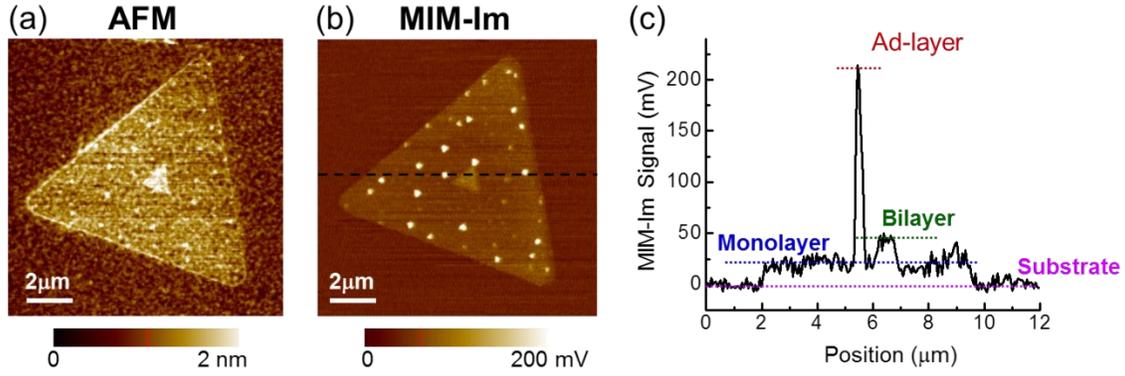

Figure S4(a, b) shows the simultaneously taken AFM and MIM data on a monolayer MoS$_2$ sample with both randomly distributed dendritic ad-layers and a big triangular island in the middle. As seen clearly in the image and the line trace in Fig. S4(c), the ad-layers, although smaller in area, show much higher MIM signals than that of the triangular bilayer island. In other words, for MoS$_2$ bilayers with good crystallinity, the conductivity is only weakly higher than that of the monolayer and is considerably lower than that of the dendritic ad-layers. Similar results have also been seen in mechanically exfoliated samples (not shown). Our main result of the high local conductivity of ad-layers is therefore nontrivial and cannot be simply attributed to the thickness effect.

## S5: Finite Element Analysis (FEA) modeling

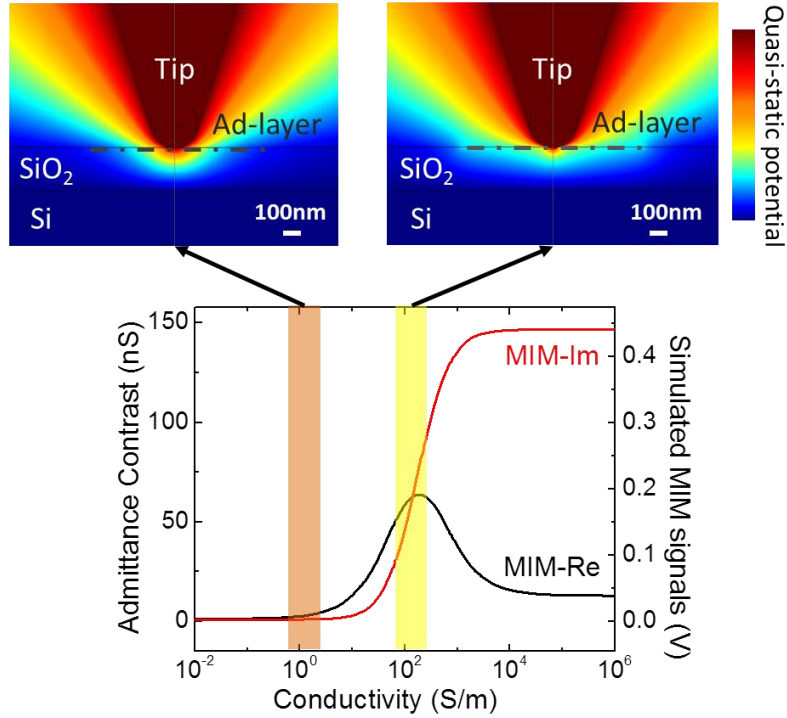

FEA simulation using COMSOL4.3 is shown in Fig. S5. Since all dimensions here are much smaller than the wavelength (30cm) at 1GHz, the interaction is in the extreme near-field regime and can be modeled as lumped elements. The $MoS_2$ conductivity is the variable to be determined in the modeling. The fixed parameters are as follows – Tip: 300nm in diameter; $SiO_2$: 285nm thick and $\varepsilon = 3.9$; heavily doped Si: $\sigma = 10^5$S/m; $MoS_2$ monolayer: 0.6nm thick and $\varepsilon = 7$ [3]; $MoS_2$ ad-layer: an extra height of 0.6nm on the monolayer, with a lateral radius of 0.7μm. Several nanometers of depletion region right underneath the tip apex are assumed to represent the non-Ohmic tip-sample contact. The software directly computes the admittance contrast with respect to the substrate as a function of the ad-layer conductivity. The MIM signals on the monolayer $MoS_2$ are barely above the noise level, corresponding to a low conductivity $\sigma_{monolayer}$ 1 ~ 10S/m here. The dendritic ad-layer, on the other hand, is much brighter in the images with roughly equal MIM-Im and MIM-Re signals. The quasi-static potential distribution at $\sigma_{ad\text{-}layer} = 10^2 \sim 10^3$ S/m shows a nearly constant potential across the ad-layer, which also explains the linear areal dependence of the MIM signals. We note that since the experimental data are quite scattered and the simulation strongly depends on the above parameters, the deduced conductivity only serves as an order of magnitude estimate.